\documentclass[twocolumn,showpacs,aps,prl,superscriptaddress]{revtex4}
\usepackage{graphicx}
\begin{document}
%
%
%
\author{Y.-M. Choi}
\affiliation{Dept. of Physics and Cond. Matter Research Inst.,
Seoul National Univ., Seoul, Korea.}
\author{D.-S. Lee}
\affiliation{Dept. of Physics and Cond. Matter Research Inst.,
Seoul National Univ., Seoul, Korea.}
\author{R. Czerw}
\affiliation{Dept. of Physics and Astronomy, Clemson University,
Clemson, SC.}
\author{P-W Chiu}
\affiliation{Max-Planck-Institut f\"{u}r Festk\"{o}rperforschung,
Stuttgart, Germany.}
\author{N. Grobert}
\affiliation{The Nanoscience and Nanotechnology Centre, CPES,
Univ. of Sussex, Brighton, England}
\author{M. Terrones}
\affiliation{IPICYT, Av. Venustiano Carranza 2425-A, San Luis
Potos\'{i} 78210, M\'{e}xico}
\author{M. Reyes-Reyes}
\affiliation{IPICYT, Av. Venustiano Carranza 2425-A, San Luis
Potos\'{i} 78210, M\'{e}xico}
\author{H. Terrones}
\affiliation{IPICYT, Av. Venustiano Carranza 2425-A, San Luis
Potos\'{i} 78210, M\'{e}xico}
\author{J-C Charlier}
\affiliation{Universit\'{e}  Catholique de Louvain, Unit\'{e} de
Physico-Chimie et de Physique des Mat\'{e}riaux, Place Croix du
Sud 1, B-1348 Louvain-la-Neuve,  Belgium}
\author{P. M. Ajayan}
\affiliation{Dept. of Materials Science and Engineering Rensselaer
Polytechnic Institute, Troy NY}
\author{S. Roth}
\affiliation{Max-Planck-Institut f\"{u}r Festk\"{o}rperforschung,
Stuttgart, Germany.}
\author{D. L. Carroll}
\affiliation{Dept. of Physics and Astronomy, Clemson University,
Clemson, SC.}
\author{Y.-W. Park}
\affiliation{Dept. of Physics and Cond. Matter Research Inst.,
Seoul National Univ., Seoul, Korea.}
\title{Nonlinear behavior in the Thermopower of Doped Carbon Nanotubes
\\Due to Strong, Localized States}
\date{\today }
\begin{abstract}
The temperature dependent thermoelectric power (TEP) of boron and
nitrogen doped multi-walled carbon nanotube mats has been measured
showing that such dopants can be used to modify the majority
conduction from p-type to n-type.  The TEP of boron doped
nanotubes is positive, indicating hole-like carriers.  In
contrast, the nitrogen doped material exhibits negative TEP over
the same temperature range, suggesting electron-like conduction.
Therefore, the TEP distinct nonlinearites are primarily due to the
formation of donor and acceptor states in the B- and N- doped
materials. The sharply varying density of states used in our model
can be directly correlated to the scanning tunneling spectroscopy
studies of these materials.
\end{abstract}
\maketitle \pagestyle{myheadings} \markright{submitted to PRL}
Interest in the electrical transport properties of both
single-walled (SWNT) and multi-walled (MWNT) carbon nanotubes
stems primarily from anticipated applications of these
low-dimensional materials in nanoelectronics \cite{dekker}.   One
expects metallic conduits along with heterojunctions, formed from
nanomaterials with different carriers (electrons vs. holes), to be
necessary. These would play an analogous role to 'bulk' doped Si
devices and metal interconnect lines.  However, the direct
substitutional doping of carbon nanotubes has proven to be quite
difficult. Their low-dimensional structure does not provide an
energetically favorable environment for most impurity atoms.
Fortunately, there are two promising exceptions, boron \cite{red}
and nitrogen \cite{terrones}, both of which seem able to reside
within the carbon lattice.

In this Letter we report on the effect of donor and acceptor
states on the thermoelectric power (TEP) of boron doped and
nitrogen doped nanotube mats.  TEP is an important and sensitive
test of the carrier sign of any material. Because the TEP is a
zero current transport coefficient, it can probe the intrinsic
conduction properties of individual nanotubes while being less
influenced by randomly entangled morphologies and imperfections of
the measured  mats as compared to standard conductivity
measurements \cite{kaiser01}.  For example, intrinsic-metallic
properties were well demonstrated with the TEP of conducting
polymer films exhibiting randomly entangled fibrillar
morphologies. However, in these systems, the electrical
conductivity always showed a semiconducting temperature dependence
due to interfibrillar junction resistances \cite{park80}.
Generally, SWNT and MWNT randomly orientated mats show a positive
and moderately large thermoelectric power (TEP) over the
temperature range of 0 to 300 K, with temperature dependencies
that approach zero as T$_{0}\rightarrow$0 \cite{bax}.  However,
earlier studies have shown that "doping" the SWNT mats by
intercalating alkali metals into the nanotube bundles could give
rise to significantly different thermoelectric properties
\cite{grig}.  Likewise, more recent work has demonstrated an
extreme sensitivity of the thermoelectric properties of SWNT mats
to oxygen contamination (or doping) \cite{brad}. In fact, this
exposure to various atmospheres has become a central issue in
understanding thermoelectric measurements in SWNT systems.  In
both cases, changes in the sign or magnitude of the TEP arise from
modifications of the density of states near the Fermi level of the
semiconducting tubes contained in the mat, and relatively little
effect is attributed to dopant interactions with the metallic
tubes in the sample.  Furthermore, these dopants were not included
within the lattice structure of the carbon nanotubes, but rather
they were attached to the outside of the cylindrical shell
\cite{grig}. Thus, they represent something other than intrinsic
behavior to this material.

The behavior of the TEP in multi-walled mats is likely to be
somewhat different from that of SWNT mats \cite{bax}.  This is
because in SWNT mats the microstructure is composed of an
assortment of semiconductors and metals, while in the MWNT mats,
the tubes are all semi-metals and small band-gap semiconductors.
In fact, many models treat the materials surprisingly the same
with a great deal of success \cite{model}. The theoretical
approach to MWNT mats adopted by most researchers has been the
inclusion of parallel metallic and semiconducting conduction paths
each broken by tube-tube contact barriers.  While this view is
clearly adequate for understanding pure mats of MWNTs,
modifications are required when addressing the TEP of doped
materials.  Here we demonstrate that while the rather large
positive TEP for pure materials might arise from the random
orientation of contact barriers within the mat and pure
semiconducting pathways, B-doped nanotube mats exhibit a positive
TEP, with a more metallic-like thermopower, and derived entirely
from the sign of the carrier in the isolated nanotubes. Further,
N-doped nanotube mats exhibit a metallic negative TEP that is
again a reflection of the isolated nanotube's behavior. Thus,
control over the nature of the carrier in MWNTs has been achieved
in direct analogy to bulk semiconductors.

\begin{figure}
\centering
\includegraphics[angle=0,width=1.0\columnwidth,clip]{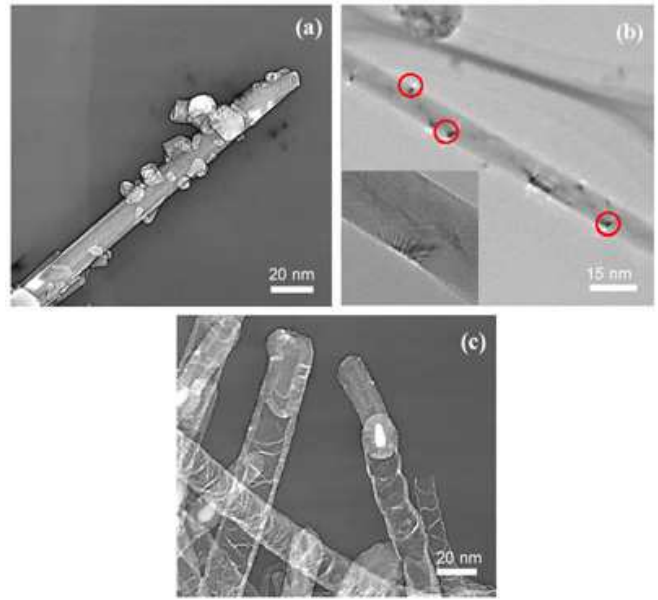}
\caption{\label{fig1} Transmission electron micrographs of the
three nanotube materials used in this study. (a) Pure MWNT
exhibiting high degree of cristallinity, note that polyhedral
particles adhere to the surface. However, we removed this
contamination after filtration techniques; (b) B-doped MWNT
showing lattice distortions within the tubes, which likely can be
attributed to the formation of local BC$_{3}$ domains within the
predominantly hexagonal carbon lattice (dark contrast regions);
(c) N-doped MWNT exhibiting a bamboo-like morphology.}
\end{figure}

For these studies, a variety of growth and processing techniques
had to be used to prepare the pure and doped materials.  The pure
carbon nanotubes were arc grown using methods described in detail
elsewhere \cite{ebb}. Transmission electron microscopy (TEM)
showed a diameter distribution to be centered around 20 nm with
tubes as small as 3 nm and as large as 40 nm. Tube lengths were
typically 1-3 $\mu$m and the primary impurities were carbonaceous
materials and polyhedral particles,  as shown in Fig.\
\ref{fig1}a. B-doped MWNTs were also grown using arc methods as
described in the literature \cite{red94}. TEM characterization
showed these materials to have typical tube diameters of 20 nm
with a range of 5 nm to 40 nm.  Selected area diffraction (SAD)
confirms that these tubes possess a predominantly zig-zag
chirality \cite{blase}.  Previously, tunneling microscopy and
spectroscopy, coupled with electron energy loss spectroscopy
(EELS), has been used to demonstrate that the boron is
incorporated into the lattice as islands of BC$_{3}$
\cite{carroll}. A visual indication of these islands could be
attributed to the presence of strained regions within the tubes
(Fig.\ \ref{fig1}b dark contrast regions). The impurities in the
growth materials were found to be polyhedral particles and small
concentrations of amorphous carbon. No catalysts are used in the
growth of either of the arc-produced materials.

Finally, the N-doped materials were grown using catalytic
decomposition of melamine-ferrocene mixtures \cite{terrones}. This
technique has only been developed recently and the tubes are well
formed with a typical diameter of around 20 nm.  Our CVD tubes
exhibit only small amounts of polyhdra particles and amorphous
material. However, much of the material had encapsulated Fe cores
at the very tip of the nanotubes that typically would extend 20 -
30 nm in length as seen in Fig.\ \ref{fig1}c. All of the
micrographs in Fig.\ \ref{fig1} show the materials ``as grown''
before filtering of the impurities and prior to the preparation of
the mats.  We note that while there are few conformational
differences between the pure and B-doped MWNTs (except chirality),
the N-doped materials exhibit a ``bamboo - like'' tube morphology.
The tubular segments within the N-doped material are approximately
100 nm in length. In the mats, the segment length is generally
longer than the average distance between tube-tube contacts.

Nanotube mats were produced by suspending each (as grown) material
into chloroform and then ultrasonically agitating until the
nanotubes were well dispersed. The suspended materials were found
to be relatively undamaged, after the extended ultrasonication,
using TEM.  Each solution was allowed to settle and then was
filtered multiple times using 0.4 mm Teflon filters.  The
remaining solution was composed of nanotubes with little amorphous
materials and few polyhedra. Finally a thick film was built-up
from each of the materials using a Teflon filter and a polyimide
mold (to insure equal dimensions in each case).  This resulted in
a random packing of nanotubes in a dense mat.  The samples were 3
mm x 5 mm x 0.025 mm in size. Several mats of each material were
made, and the measurements performed several times on each to
insure reproducibility.  In order to carry out thermoelectric
power measurements, the mat samples were supported on a Teflon
substrate and mounted on top of two coppers blocks.  Silver paste
was used for the electrical contacts.  Chromel-Constantan
thermocouples were attached to the back of the copper blocks using
GE 7031 varnish.  Techniques for the TEP measurement of carbon
nanotubes are described with more detail in the literature
\cite{kim}.

\begin{figure}
\centering
\includegraphics[angle=0,width=1.0\columnwidth,clip]{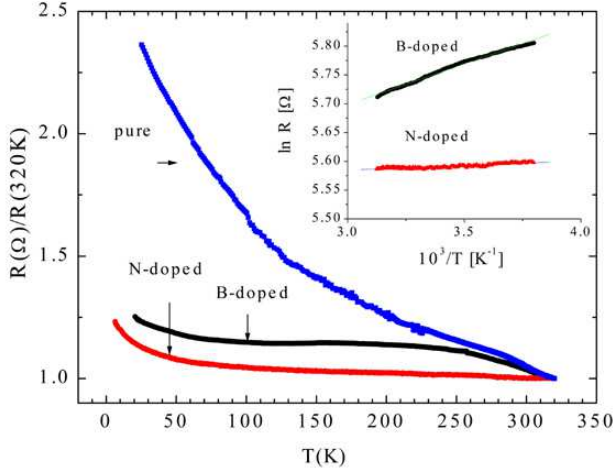}
\caption{\label{fig2} Two-probe measurements of the temperature
dependence of R for pristine, b-doped, and n-doped nanotube mats.
All three mats exhibited non-metallic behavior over the entire
temperature range.}
\end{figure}

Fig.\ \ref{fig2} shows a two-probe measurement of the temperature
dependence of R, normalized to 320 K,  after degassing in vacuum
at 320 K for three days. All three mats exhibited non-metallic
behavior over the entire temperature range (dR/dT$<$0).  The
primary effect of the lattice doping was to reduce the temperature
dependence of R. The ratio R$_{20}$/R$_{320}$  was reduced from
$\sim$2.4 in the pristine case to $\sim$1.25 and $\sim$1.14 in the
boron and nitrogen cases, respectively. Similar reductions in the
temperature dependence have been reported for I-doped and Cs-doped
SWNT mats \cite{grig}. Fig.\ \ref{fig2} inset shows linear fits to
an Arrhenius plot of R versus the absolute temperature T. For
sufficiently high temperatures, where intrinsic conductivity
dominates, one can expect resistivity to follow an Arrhenius type
equation of the form:
\begin{equation}
\rho=\rho_{0}exp(E_{rho}/kT),
\end{equation}
where E$_{\rho}$ is the activiation energy, $ \rho_{0}$ is a
constant, and k is Boltzman's constant.  For the B-doped mat, the
acivation energy (12.2 meV) was found to be larger than the
activation energy of the N-doped mats (1.4 meV), but smaller than
the activation energy reported for individual B-doped tubes (55-70
meV) \cite{wei}.
\begin{figure}
\centering
\includegraphics[angle=0,width=1.0\columnwidth,clip]{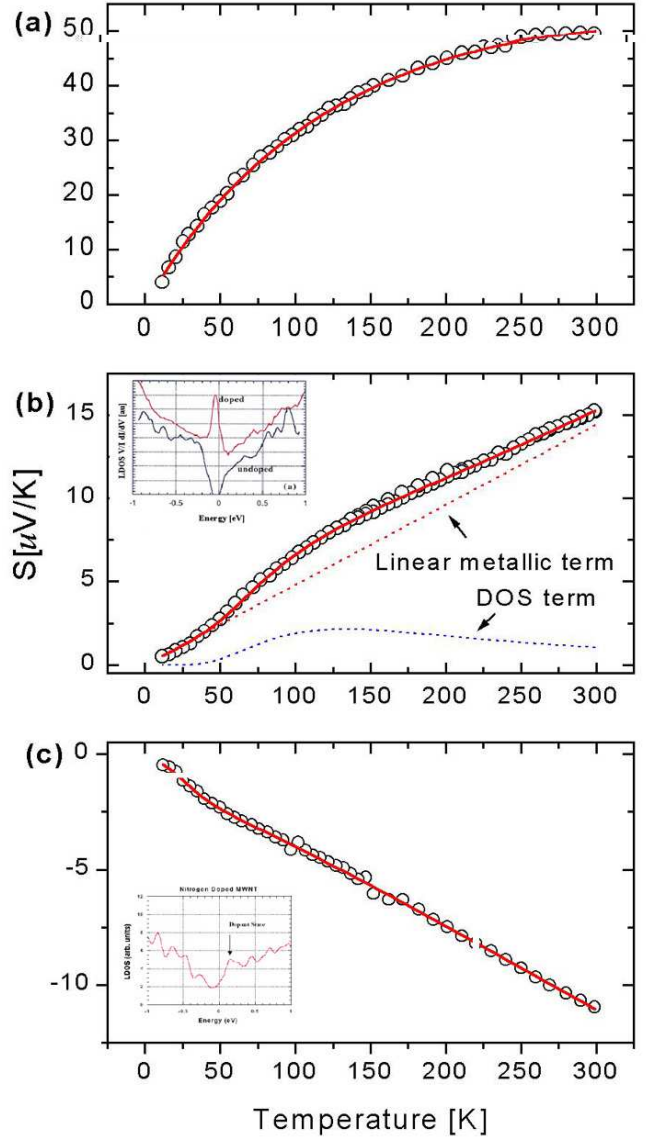}
\caption{\label{fig3} Thermopower of pure and doped MWNT mats. (a)
TEP of pure MWNT mat fit by heterogeneous model (solid line); (b)
TEP of B-doped MWNT mat and fit (solid line) indicating positve
(hole) carriers. The dotted lines show the effect of the linear
metallic term (straight line) and the DOS term.  The DOS peak
results in a broader peak in S; (c) TEP of N-doped MWNT mat
indicating negative (electron) carriers and fit (solid line).}
\end{figure}

The TEP of the pure mats is shown in Fig.\ \ref{fig3}a, and the
solid line shows the fit using a heterogenous model.  Each
percolating pathway is assumed to have a randomly oriented set of
barriers due to tube-tube contacts, tube defects, etc.,  It is
therefore expected, for such a model, that the TEP will be
consistent with the sum of a linear metallic term and an
exponentially weighted T$^{1/2}$ variable range hopping term that
reflects the ``freezing-out'' of semiconducting contributions at
low temperatures \cite{kaiser99} given by the following equation:
\begin{equation}
S=bT+qT^{1/2}exp[-(T_{0}/T)^{1/4}],
\end{equation}
where bT is is the metallic term and q is a constant. However,
although this model gives a reasonable fit of the pure data, the
TEP is the sum of a very large and un-physical positive metallic
term (135 $\mu$V/K at 300 K), and a correspondingly large negative
semiconducting term. The large positive TEP is due in part to
oxygen doping of the pure mats.   The TEP of the pure MWNT mats in
our study was found to decrease by 10.9 $\mu$V/K at 320 K when the
sample was exposed to vacuum for 94 hours.  This is in constrast
to the boron and nitrogen doped MWNT mats which showed no
variation in TEP after similar time exposure to vacuum.

In the case of the doped materials, we suspect nonlinearities are
introduced into the linear metallic thermopower by sharply varying
and localized states near the fermi level due to the addition of
boron or nitrogen into the lattice. Such states have been
predicted and subsequently verified experimentally by scanning
tunneling spectroscopy (STS) \cite{carroll, czerw}.  Taking the
standard expression for thermopower \cite{dugdale}:
\begin{equation}
S=\frac{1}{eT\sigma}\int\sigma(E)(E-\mu)\frac{df}{dE}dE,
\end{equation}
where $\sigma$ is the conductivity, $\mu$ is the chemical
potential (assumed to be constant with temperature), $\sigma$(E)
the partial conductivity at energy E, and $f$ the Fermi function,
we approximate the partial conductivity function by a delta
function due to the sharply varying density of states near the
Fermi level.  This gives the following expression for the
thermopower \cite{kaiser99}:
\begin{equation}
S=bT+\frac{qT_{p}}{e\sigma
T^{2}}\frac{exp(T_{p}/T)}{[exp(T_{p}/T)+1]^{2}}.
\end{equation}
In this expression, bT is the linear metallic term, q is a
constant, and T$_{p}$=(E-$\mu$)/k where E is the energy at which
the peak occurs and k is Boltzmann's constant.  As can be seen in
Fig.\ \ref{fig3}b and Fig.\ \ref{fig3}c, Eq. (4) gives a good fit
with E=27 meV and 11 meV for the boron and nitrogen samples,
respectively.  From the LDOS (Fig.\ \ref{fig3}b inset), the sign
of the TEP for the B-doped materials is not surprising since these
nanotubes have strong acceptor states and hole-like conductivity
should be expected.  In contrast, the N-doped nanotubes show a
negative TEP with a similar functional form to that of the B-doped
materials.  From the large donor state in the LDOS (Fig.\
\ref{fig3}c inset), it is clear that these materials should be
electron rich and thus the sign of the TEP is expected.  The shape
of the TEP for the doped materials is also very similar to that of
the TEP curve for iodine-doped SWNT bundles \cite{grig}.  Since
chloroform was used to suspend the tubes during sonication, and
chloroform has been shown to result in C-Cl bonds on nanotube
surfaces or to react with the amorphous carbon in bundles
\cite{lee01}, it may be argued that the effects are due to doping
by Cl$^{-}$.  However, in the previous study, extremely long
sonications and reflux times in the presence of a catalyst were
used to form the bonds.  It is doubtful that the shorter
sonication times in our samples ($<$ 30 min) could result in
significant doping by Cl$^{-}$. Also, the sign of the carrier of
the n-doped mats is clearly negative.

In summary, the thermoelectric power in mats of doped MWNT's
reflects the sign of the majority carrier of the individual
nanotubes. Further, the behavior of the TEP has been directly
correlated with the density of electronic states near Fermi level
as determined by STS and gives a positive TEP when acceptor states
are present as well as a negative TEP when donor states are
present. The TEP of pure MWNT mats has shown some sensitivity to
air exposure, while the doped mats are effectively insensitive to
``oxygen doping.'' The TEP is strongly dominated by the boron and
nitrogen doping and the resulting donor and acceptor states near
the Fermi level.

The authors gratefully acknowledge support from: KISTEP
98-I-01-04-A-026, MOST (Korea), AFOSR F49620-99-1-0173 (U.S.), DFG
(Germany). MT and HT are grateful to CONACYT-Mexico grant W-8001
millennium initiative. NG acknowledges the Royal Society for
financial support, and PMA thanks Philip Morris U.S.A. for their
financial support.


\begin{thebibliography}{}
%
\bibitem{dekker}See for example: C. Dekker, Phys. Today \textbf{52}, No. 5, 22 (1999).
%
\bibitem{red}
P. Redlich, J. Loeffler, P.M. Ajayan, J. Bill, F. Aldinger, and M. R\"{u}hle,
Chem Phys. Lett.\textbf{260}, 465 (1994).
%
\bibitem{terrones}
M. Terrones \textit{et al}., Appl. Phys. Lett. \textbf{75}, 3932 (1999); W.Q. Han
\textit{et al}., Appl. Phys. Lett. \textbf{77}, 1807 (2000).
%
\bibitem{kaiser01}
A. B. Kaiser, Adv. Mater. \textbf{13}, no. 12-13, 927 (2001).
%
\bibitem{park80}
Y. W. Park, A. J. Heeger, M. A. Druy, and A. G. MacDiarmid, J. Chem. Phys. \textbf{73},
946 (1980).
%
\bibitem{bax}
M. Baxendale, K.G. Lim, and G. A. J. Amaratunga, Phys. Rev. B, \textbf{61}, 12705 (2000).
%
\bibitem{grig}
L. Grigorian, G. U. Sumanasekera, A. L. Loper, S. Fang, J.L. Allen, and P.C. Eklund,
Phys. Rev. B \textbf{58}, R4195 (1998).
%
\bibitem{brad}
K. Bradley, S.-H. Jhi, P. G. Collins, J. Hone, M. L. Cohen, S. G.
Louie, and A. Zettl, Phys. Rev. Lett. \textbf{85}, 4361 (2000).
%
\bibitem{model}
Compare for example, the model of Ref. 6 with that of A.B. Kaiser
\textit{et al}., Phys. Rev. B \textbf{57}, 1418 (1998).
%
\bibitem{ebb}
T. W. Ebbesen and P. M. Ajayan Nature \textbf{358}, 220 (1992).
%
\bibitem{red94}
P. Redlich \textit{et al}., Chem Phys. Lett. \textbf{260}, 465 (1994).
%
\bibitem{blase}
X. Blase \textit{et al}., Phys. Rev. Lett. \textbf{83}, 5078
(1999).
%
\bibitem{carroll}
D. L. Carroll, P. Redlich, X. Blase, J. C. Charlier, S. Curran,
P.M. Ajayan, S. Roth, and M. R\"{u}hle, Phys. Rev. Lett.
\textbf{81}, 2332 (1998).
%
\bibitem{kim}
Y.-W. Park, Synth Met. \textbf{45}, 173 (1991).
%
\bibitem{wei}
B. Wei, R. Spolenak, P. K. Redlich, M. Ruhle, and E. Arzt, Appl.
Phys. Lett. \textbf{74}, 3149 (1999).
%
\bibitem{kaiser99}
A. B. Kaiser, Y. W. Park, G. T. Kim, E. S. Choi, G. D\"{u}sberg,
and S. Roth, Synth. Met. \textbf{103}, 2547 (1999).
%
\bibitem{czerw}
R. Czerw \textit{et al}., Nano Lett. \textbf{1}, 457 (2001).
%
\bibitem{dugdale}
J. S. Dugdale, The Electrical Properties of Metals and Alloys,
Edward Arnold, Paris, (1977).
%
\bibitem{lee01}
W. H. Lee, S. J. Kim, W. J. Lee, J. G. Lee, R. C. Haddon, and P.
J. Reucroft, Appl. Surf. Sci. \textbf{181}, 121 (2001).
%
%
\end{thebibliography}
\end{document}